\documentclass[useAMS,usenatbib]{mn2e}

\usepackage{amsfonts,amsmath,amssymb,mathrsfs}
\usepackage{graphicx,epsf,epsfig}
\usepackage{bm}
\usepackage{longtable}
\usepackage[usenames,dvipsnames]{xcolor}
\usepackage{multirow}

\usepackage{graphicx}
\usepackage{times}
\usepackage{color}
\usepackage{breakurl}
\usepackage{color}
\usepackage{mathrsfs}

\usepackage{hyperref}
\hypersetup{
  colorlinks=true,        
  linkcolor=blue,         
  citecolor=cyan,         
}

\def\be{\begin{equation}}

\newcommand{\cf}{cf.,~}
\newcommand{\ie}{i.e.,~}
\newcommand{\eg}{e.g.,~}

\title[Maximum mass, moment of inertia, compactness]{Maximum mass, moment of inertia and compactness of relativistic
  stars}

\author[Breu and Rezzolla]{Cosima Breu and Luciano Rezzolla \\ \\
Institute for Theoretical Physics, Max-von-Laue-Str. 1, 
60438 Frankfurt, Germany \\
Frankfurt Institute for Advanced Studies, Ruth-Moufang-Str. 1, 
60438 Frankfurt, Germany}

\begin{document}
\date{\today}
\maketitle

\pagerange{\pageref{firstpage}--\pageref{lastpage}} 

\maketitle

\label{firstpage}

\begin{abstract}
A number of recent works have highlighted that it is possible to express
the properties of general-relativistic stellar equilibrium configurations
in terms of functions that do not depend on the specific equation of
state employed to describe matter at nuclear densities. These functions
are normally referred to as ``universal relations'' and have been found
to apply, within limits, both to static or stationary isolated stars, as
well as to fully dynamical and merging binary systems.
Further extending the idea that universal relations can be valid also
away from stability, we show that a universal relation is exhibited also
by equilibrium solutions that are not stable. In particular, the mass of
rotating configurations on the turning-point line shows a universal
behaviour when expressed in terms of the normalized Keplerian angular
momentum. In turn, this allows us to compute the maximum mass allowed by
uniform rotation, $M_{\rm max}$, simply in terms of the maximum mass of
the nonrotating configuration, $M_{_{\rm TOV}}$, finding that $M_{\rm
  max} \simeq \left(1.203 \pm 0.022\right) M_{_{\rm TOV}}$ for all the
equations of state we have considered.
We further introduce an improvement to previouly published universal
relations by Lattimer \& Schutz between the dimensionless moment of
inertia and the stellar compactness, which could provide an accurate
tool to constrain the equation of state of nuclear matter when
measurements of the moment of inertia become available.


\end{abstract}


\section{Introduction}
\label{sec:introduction}

The behaviour of nuclear matter at the extreme densities reached in
neutron-star cores is determined by the properties of fundamental
interactions in regimes that are still poorly known and that are not
accessible in experiments on Earth. A broad variety of different models
for supernuclear matter, and thus for the inner structure of neutron
stars, has been proposed. Although neutron-star properties depend
sensitively on the equation of state (EOS), approximately ``universal''
relations between several dimensionless quantities have been found during
the last years. The ``universal'' character in these relations comes from
their weak dependence on the EOS and this allows one to use them to
constrain quantities that are difficult to access experimentally, such as
the neutron star radius.

Already in 1994, \citet{Ravenhall94} highlighted an apparently universal
relation between the normalized moment of inertia $I/(MR^{2})$ and the
stellar compactness $M/R$ in the case of EOSs without an extreme
softening at supernuclear densities. This relation was later refined by
\citet{Lattimer01} and \citet{Bejger02}, and then employed by
\citet{Lattimer2005b} to point out that using such an empirical relation
it is possible to estimate the radius of a neutron star via the combined
measurement of the mass and moment of inertia of a pulsar in a binary
system. Additional evidence that EOS-independent relations could be found
among quantities characterizing compact stars was also pointed out by
\citet{Andersson1998b} [and later on revisited by \citet{Benhar:2004xg}],
who showed that a tight correlation exists between the frequency of the
fundamental mode of oscillation and the stellar average density. More
recently, however, \citet{Yagi2013a} have found that suitably normalized
expressions for the moment of inertia $I$, the quadrupole moment $Q$ and
the tidal Love number $\lambda$ are related by functions independent of
the EOS to within $\sim 1\%$ in the slow-rotation approximation and
assuming small tidal deformations [see also the related and earlier work
  by \citet{Urbanec2013}; \citet{Baubock2013}]. These relations are particularly useful as
they may help remove some degeneracies as those appearing, for instance,
in the modelling of the gravitational-wave signal from inspiralling
binaries [see \citet{Maselli2013}; \citet{Yagi2013b} for a
  discussion].

The universality, however, should be meant as approximate and is indeed
preserved only within large but defined regimes, such as the
slow-rotation approximation or when the magnetic fields are not
particularly strong. More specifically, it has been shown by
\citet{Doneva2014a} that the universality in the relation between $I$ and
$Q$ breaks down in the case of rapid rotation for sequences with constant
spin frequency. This result was partially revised by
\citet{Chakrabarti2014}, who have shown that the $I$--$Q$ universality is
partly recovered if the rotation is characterized not by the spin
frequency, but by the dimensionless angular momentum $j:=J/M^{2}$, where
$J$ and $M$ are the angular momentum and mass of the star, respectively
[see also \citet{Pappas2014}]. At the same time, \citet{Haskell2014} have
shown that the universality between $I$ and $Q$ is lost for stars with
long spin periods, \ie $P \gtrsim 10\,$s, and strong magnetic fields, \ie
$B \gtrsim 10^{12}\,$G. The reason behind this behaviour is rather
simple: the anisotropic stresses introduced by a magnetic field can break
the overall spherical symmetry present for stars that are nonrotating or
in slow rotation. In turn, this affects the universal behaviour, which is
based on the balance between gravity and the behaviour of fluids in
strong gravitational fields.
 
Over the last couple of years, this area of research has been extremely
active, with investigations that have considered universality also with
higher multipoles \citep{Chatziioannou2014, Yagi2014, Pappas2014, Stein2014} or that have sought universality in response to alternative
theories of gravity \citep{Doneva2014b, Kleihaus2014, Pani2014}. A large
bulk of work has also tried to provide a phenomenological justification
for the existence of such a behaviour. For instance, it has been
suggested that $I$, $\lambda$ and $Q$ are determined mostly by the
behaviour of matter at comparatively low rest-mass densities, where realistic
EOSs do not differ significantly because they are better constrained by
nuclear-physics experiments \citep{Yagi2013b}. Alternatively, it has been
suggested that the multipole moments approach the limiting values of a
black hole towards high compactness, which implies that approximate
no-hair-like relations exist also for non-vacuum spacetimes
\citep{Yagi2014b}. Finally, it has been proposed that nuclear-physics
EOSs are stiff enough that the nuclear matter can be modelled with an
incompressible EOS. In this interpretation, low-mass stars, which are
composed mainly of soft matter at low densities, would depend more
sensitively on the underlying EOS, while the universality is much
stronger towards higher compactness, where matter approaches the limit of
an incompressible fluid \citep{Martinon2014, Sham2014b}.

In this paper, we take a slightly different view and reconsider
well-known universal relations to devise an effective tool to constrain
the radius of a compact star from the combined knowledge of the mass and
moment of inertia in a binary system containing a pulsar. More
specifically, we first show that a universal relation holds for the
maximum mass of uniformly rotating stars when expressed in terms of the
dimensionless and normalized angular momentum. This relation extends and
provides a natural explanation for the evidence brought about by
\citet{Lasota1996} of a proportionality between the maximum mass allowed
by uniform rotation and the maximum mass of the corresponding nonrotating
configuration. Finally, we show that the dimensionless moment of inertia
$\bar{I}:=I/M^3$ for slowly rotating compact stars correlates tightly and
universally with the compactness $\mathscr{C} := M/R$, where $R$ is the
radius of the star. Next, we provide an analytical expression for such a
correlation and show that it improves on previous expressions, yielding
relative errors in the estimate of the radius that are $\lesssim 7\%$ for
a large range of masses and moment of inertia. We should note that
universal relations $\bar{I}$--$\lambda$ and $\lambda$--$\mathscr{C}$
have been shown to hold by \citet{Yagi2013b} and \citet{Maselli2013},
respectively. Hence, it is not surprising that a universal relation
$\bar{I}$--$\mathscr{C}$ also holds. However, to the best of our
knowledge, this is the first time that such a relation is discussed in
detail and that its implications are investigated in terms of
astrophysical measurements.

The plan of the paper is as follows. In Section \ref{sec:setup}, we
briefly review the mathematical and numerical setup used for the
calculation of our equilibrium stellar models (from nonrotating
configurations to rapidly rotating ones). Section \ref{sec:maximumass} is
dedicated to our results and in particular to the new universal relation
between the maximum mass and the normalized angular momentum. Section
\ref{sec:results} is instead dedicated to the comparison between our
expression for the dimensionless moment of inertia and previous results
in the literature, while in Section \ref{sec:applications} we discuss how
the new relation can be used to deduce the radius of a compact star once
a measurement is made of its mass and moment of inertia. Finally, Section
\ref{sec:conclusions} contains our conclusions and prospects for future
work; the Appendix provides the derivation of an analytic expression to
evaluate the relative error in the estimate.

\section{Mathematical and numerical setup}
\label{sec:setup}

We have constructed a large number of equilibrium models of compact stars
that are either nonrotating, slowly or rapidly rotating. In the first two
cases, solutions can be obtained after the integration of ordinary
differential equations in spherical symmetry, while in the third case we
have made use of a two-dimensional numerical code solving elliptic
partial differential equations.

More specifically, we consider the stellar matter as a perfect fluid with
energy-momentum tensor \citep{Rezzolla_book:2013} 
\begin{equation}
T^{\mu\nu}=(e+p)u^{\mu}u^{\nu}+pg^{\mu\nu}\,,
\end{equation}
where $u^{\mu}$ is the fluid 4-velocity, $g^{\mu \nu}$ the four-metric,
while $e$ and $p$ are the fluid energy density and pressure,
respectively. In the case of non-rotating stars, we take a spherically
symmetric metric 
\begin{align} 
\label{eq:nonrot}
\mathrm{d}s^{2}=-\mathrm{e}^{2\phi(r)}\mathrm{d}t^{2}+\mathrm{e}^{2\lambda(r)}\mathrm{d}r^{2}
+ r^{2} \left(\mathrm{d}\theta^{2}+\sin{\theta}^{2} \mathrm{d}\varphi^2\right)\,,
\end{align}
where $\phi(r)$ and $\lambda(r)$ are metric functions of the radial
coordinate $r$ only. Equilibrium models are then found as solutions of
the Tolman--Oppenheimer--Volkoff (TOV) equations
\begin{align}
\label{eq:M}
\frac{\mathrm{d}m(r)}{\mathrm{d}r}&=4\upi r^{2}e\,,\\
\label{eq:P}
\frac{\mathrm{d}p(r)}{\mathrm{d}r}&=\frac{(e+p)(m+4\upi r^{3}p)}{r(r-2m)}\,,\\
\label{eq:nu}
\frac{\mathrm{d}\phi(r)}{\mathrm{d}r}&=\frac{2}{(e+p)}\frac{\mathrm{d}p}{\mathrm{d}r}\,,
\end{align}
where the function $m(r)$ is defined as
\begin{equation}
\label{eq:m_of_r}
m(r) := \frac{1}{2} r(1-\mathrm{e}^{-2\lambda})\,.
\end{equation} 

These TOV equations need to be supplemented with an EOS providing a
relation between different thermodynamic quantities, and we have used 15
nuclear-physics EOSs in tabulated form. For all of them, beta equilibrium
and zero temperature were assumed, so that the EOS reduces to a relation
between the pressure and the rest-mass density (or the energy
density). For more complex EOSs in which a temperature dependence is
available, we have used the slice at the lowest temperature. In our
analysis we have considered 28 several different theoretical approaches
to the EOS and, in particular: the nuclear many-body theory [APR4,
  \citet{Akmal1998a}; WFF1, WFF2, \citet{Wiringa88}], the
non-relativistic Skyrme mean-field model [Rs, SK255, SK272, Ska, Skb,
  SkI2-SkI6, SkMp, \citet{Gulminelli2015}, SLY2, SLY4, SLY9,
  \citet{Douchin01,Gulminelli2015}], the mean-field theory approach [EOS
  L, \citet{Pandharipande75}; HS DD2, HS NL3, HS TM1, HS TMA, SFHo, SFHx,
  \citet{Gaitanos2004, Hempel2010, Typel2010}; GShen-NL3,
  \citet{ShenG2010}, the Walecka model [EOS-O, \citet{Arnett77}] and the
  liquid-drop model [LS220; \citet{Lattimer91}]. All of these models are
  able to support a neutron star with a maximum mass of at least 2.0
  $\mathrm{M}_{\bigodot}$ and are therefore compatible with the discovery of neutron
  stars with masses of about 2 $\mathrm{M}_{\bigodot}$ \citep{Demorest2010,
    Antoniadis2013}. At the same time, we note that because strong phase
  transitions above nuclear saturation density can affect the radii of
  low-mass neutron stars, determining whether a phase transition occurs
  (and if so at which rest-mass density) represents an important
  consideration not contemplated in these EOSs, but that will be the
  focus of future work.

It is possible to extend the validity of the TOV solutions by considering
stellar models in the slow-rotation approximation
\citep{Hartle67}\footnote{The slow-rotation approximation normally refers
  to treatments that are truncated at first order in the spin frequency
  $\Omega$. However, already in \citet{Hartle67}, a full second-order
  treatment was developed, and subsequently applied by \citet{Hartle68}
  for a systematic study of rotating relativistic stars.}. In this case,
spherical symmetry is still preserved, but rotational corrections do
appear at first order in the stellar angular velocity $\Omega$, and the
line element \eqref{eq:nonrot} is replaced by its slow-rotation
counterpart \citep{Hartle67}
\begin{align} 
\label{eq:slowrot}
\mathrm{d}s^{2}=&-\mathrm{e}^{2\phi(r)}\mathrm{d}t^{2}+\mathrm{e}^{2\lambda(r)}\mathrm{d}r^{2}\nonumber
\\& + r^{2}[\mathrm{d}\theta^{2}+\sin{\theta}^{2}(\mathrm{d}\varphi-(\omega(r,\theta) +
  \mathcal{O}(\Omega^{3}))\mathrm{d}t)^{2}]\,,
\end{align}
where $\phi(r)$, $\lambda(r)$ are still functions of the radial
coordinate only and $\omega(r,\theta)$ represents the angular velocity of
the inertial frames dragged by the stellar rotation. The set of equations
(\ref{eq:M})--(\ref{eq:nu}) needs then to be augmented with a
differential equation for the relative angular velocity
\begin{equation}
\label{eq:omega}
\frac{1}{r^{4}}\frac{\mathrm{d}}{\mathrm{d}r}\left(r^{4}j\frac{\mathrm{d}\bar{\omega}}{\mathrm{d}r}\right) +
\frac{4}{r}\frac{\mathrm{d}j}{\mathrm{d}r}\bar{\omega}=0\,,
\end{equation}
where $\bar{\omega} := \Omega-\omega(r)$ is the difference between the
angular velocity $\omega$ acquired by an observer falling freely from
infinity and the angular velocity of a fluid element measured by an
observer at rest at some point in the fluid. 

A numerical solution to equations (\ref{eq:M})--(\ref{eq:omega}) can be
obtained by integrating them outwards from the stellar centre using, for
instance, a fourth-order Runge-Kutta algorithm. At the stellar surface,
$\bar{\omega}$ must be matched to the analytic exterior solution which
has the form
\begin{align}
\bar{\omega}=\Omega-\frac{2J}{r^{3}}\,,
\end{align}
where $J$ is the total angular momentum of the star. The integration,
which can be started with an arbitrary value for $\bar{\omega}$, can then
be adjusted so as to match the boundary condition. Once a slowly rotating
solution with angular momentum $J$ and spin frequency $\Omega$ has been
found, the corresponding moment of inertia can then be computed as
\begin{equation}
I :=\frac{J}{\Omega}=
\frac{R^{4}}{6\Omega}\left(\frac{\mathrm{d}\bar{\omega}}{\mathrm{d}r}\right)_{R}\,,
\end{equation}
where the derivative in the last equality is meant to be taken at the
stellar surface. Note that at this order the angular momentum $J$ depends
linearly on the angular velocity $\Omega$, so that the moment of inertia
does not depend on $\Omega$. Quite generically, the moment of inertia
increases almost linearly with the mass for sufficiently small masses;
however, as the maximum mass is approached, it decreases rapidly. This
behaviour is due to the fact that $I$ depends quadratically on the
radius, which decreases significantly near the maximum mass. 

We should note that treatments higher than the first-order in $\Omega$
have been also presented, in \citet{Hartle67, Hartle73}, where even the third-order terms were included [see
  also \citet{Benhar2005}]. These treatments, however, are still not very
accurate for rapidly rotating models of compact stars. Hence, we have
relied on fully numerical solutions constructed using the open-source
code \textsc{RNS}, which solves the Einstein equations in axisymmetry
and in the conformally flat approximation \citep{Stergioulas95}. In this
case, the spacetime is assumed to be stationary and axisymmetric, and can
be described by a metric of the form
\begin{equation}
ds^{2}=-\mathrm{e}^{\gamma+\varrho}dt^{2}+\mathrm{e}^{2\alpha}(dr^{2} +
r^{2}d\theta^{2})+\mathrm{e}^{\gamma-\varrho}r^{2}\sin{\theta}^{2}
(d\varphi-\omega dt)^{2}\,, 
\end{equation}
where the metric potentials $\gamma$, $\varrho$, $\alpha$ and $\omega$
are functions of the quasi-isotropic coordinates $r$ and
$\theta$. Stellar models can then be computed along sequences in which
the central energy density and the axis ratio (or spin frequency) is
varied \citep{Stergioulas95}.

\section{Universality away from stability} 
\label{sec:maximumass}

With the exception of those works that have looked at the existence of
universality in merging systems of neutron-star binaries \citep{Read2013,
  Maselli2013, Takami:2014, Takami2015, Bernuzzi2014}, all of the work
done so far on universal relations in isolated relativistic stars has
concentrated on \emph{stable} equilibrium configurations. This is of
course the most natural part of the space of parameters that is worth
investigating. Yet, it is interesting to assess whether universality is
present also for stellar models that are either marginally stable or that
actually represent unstable equilibrium models. To this scope, we have
investigated whether a universal behaviour can be found also for models
whose gravitational mass is at the critical-mass limit or reasonably
close to it.

\begin{figure*}
\begin{center} 
\includegraphics[width=0.49\textwidth]{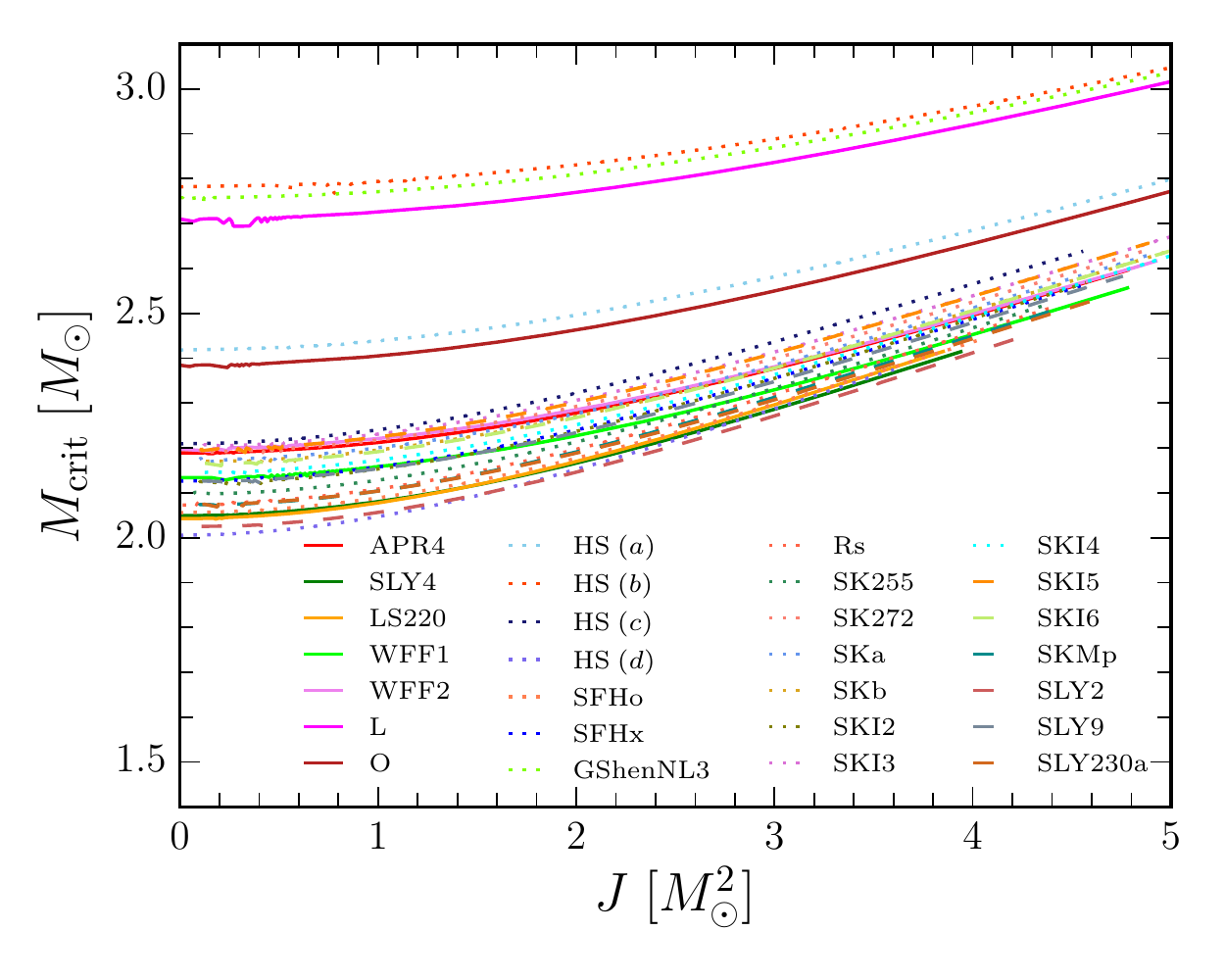}
\includegraphics[width=0.49\textwidth]{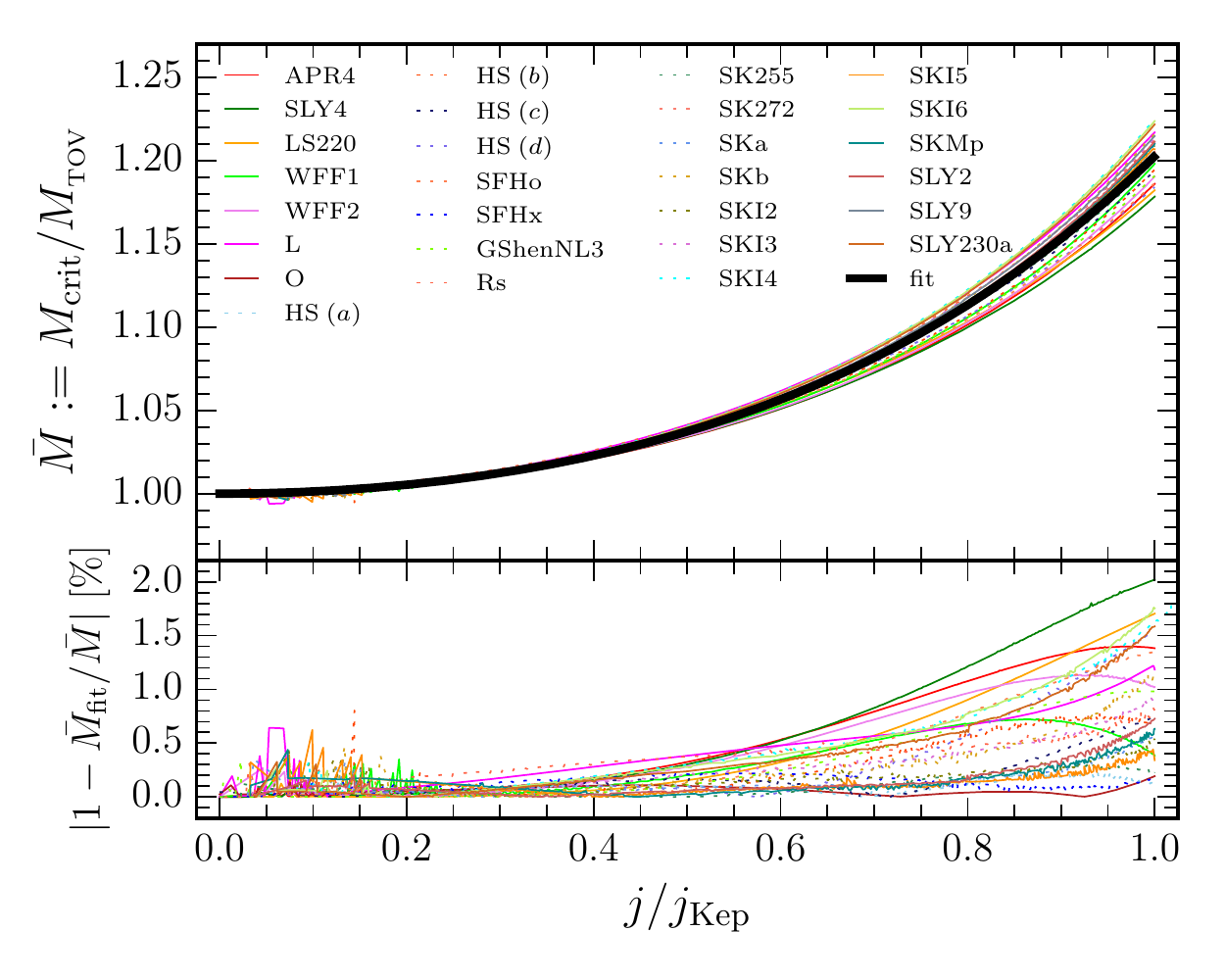}
\end{center}
\caption{\textit{Left panel:} ``Critical'' masses, \ie masses of stellar
  models along the turning-point line, shown as a function of the
  corresponding angular momentum $J$ and for a variety of EOSs. The EOSs
  $\mathrm{HS}\ (a)$--$\mathrm{HS}\ (d)$ refer to
  $\mathrm{HS}\ \mathrm{DD2}$, $\mathrm{HS}\ \mathrm{NL3}$,
  $\mathrm{HS}\ \mathrm{TM1}$, and $\mathrm{HS}\ \mathrm{TMa}$,
  respectively. \textit{Right panel:} The same data as in the left panel
  but when expressed in terms of dimensionless and normalised quantities:
  $M_{\mathrm{crit}}/M_{_{\rm TOV}}$ and $j/j_{\mathrm{Kep}}$. Shown with
  a black solid line is the polynomial fit (\ref{eq:Mcrit}), while the
  lower panel shows the relative deviation of the numerical data from the
  fitting function.}
\label{fig:critical}
\end{figure*} 

We recall that \citet{Friedman88} proved that a sequence of uniformly
rotating barotropic stars is secularly unstable on one side of a turning
point, \ie when it is an extremum of mass along a sequence of constant
angular momentum, or an extremum of angular momentum along a sequence of
constant rest-mass. Furthermore, arguing that viscosity would lead to
uniform rotation, they concluded that the turning point should identify
the onset of secular instability. While for nonrotating stars the turning
point does coincides with the secular-instability point (and with the
dynamical-instability point for a barotropic star if the perturbation
satisfies the same equation of state of the equilibrium model), for
rotating stars it is only a sufficient condition for a secular
instability, although it is commonly used to find a dynamical instability
in simulations~\citep{Baiotti04, Radice:10}. More recently,
\citet{Takami:2011} have computed the neutral-stability line for a large
class of stellar models, \ie the set of stellar models whose $F$-mode
frequency is vanishingly small; such a neutral-stability point in a
nonrotating star marks the dynamical-stability limit. The
latter coincides with the turning-point line of~\citet{Friedman88} for
nonrotating stars, but differs from it as the angular momentum is
increased, being located at smaller central rest-mass densities as the
angular momentum is increased. Stated differently, the results of
\citet{Takami:2011} have shown that equilibrium stellar models on the
turning-point line are effectively dynamically \emph{unstable}. This
result, which was also confirmed by numerical simulations, does not
contradict turning-point criterion since the latter is only a sufficient
condition for secular instability [see also the discussion by
  \citet{Schiffrin2014}].

Determining the critical mass in a way that is independent of the EOS is
critical in many astrophysical scenarios, starting from those that want
to associate Fast Radio Bursts \citep{Thornton2013} to a ``blitzar'' and
hence to the collapse of a supramassive neutron star \citep{Falcke2013},
over to those that apply this scenario to the merger of binary
neutron-star systems \citep{Zhang2014} and are thus interested in the
survival time of the merger to extract information on the EOS
\citep{Lasky2013}, or to those scenarios in which the late collapse of
the binary merger product can be used to explain the extended X-ray
emission in short gamma-ray burst \citep{Zhang2001, Rezzolla2014b,
  Ciolfi2014}.

We should note that this is not the first time that ``universal''
relations in the critical mass are considered. Indeed, \citet{Lasota1996}
have already discussed that there is a close correlation between the
``mass-shedding'' (or Keplerian) frequency $\Omega_{\rm Kep}$ and the
mass and radius of the maximum mass configuration in the limit of no
rotation, \ie $M_{_{\rm TOV}}$ and $R_{_{\rm TOV}}$. Because larger
masses can support larger Keplerian frequencies, the maximum critical
mass is defined as the largest mass for stellar models on the critical
(turning-point) line. After considering a large set of EOSs,
\citet{Lasota1996} have shown that there exists a universal
proportionality between the radii and masses of maximally rotating and of
static configurations\footnote{Interestingly, the concept of universality
  is already pointed out clearly in \citet{Lasota1996}, although it
  refers to a single value of the universal relation (\ie the maximum
  mass) and not to the whole functional behaviour.}.

In view of the simplicity of computing stellar models on the turning-point
line and given that these models are very close to those on the
neutral-stability line, we have computed the maximum masses of models
along constant angular momentum-sequences, \ie such that $\left({\partial
  M}/{\partial \rho_c}\right)_{J}=0$, where $\rho_c$ is the central
rest-mass density. Although these models are strictly speaking unstable,
for simplicity we have dubbed them ``critical masses'', $M_{\rm
  crit}$. The values of these masses as a function of the angular
momentum $J$ are reported in the left panel of of
Fig. \ref{fig:critical}, which shows that a large variance exists with
the EOS when the data is reported in this way\footnote{The data for large
  masses and small values of the angular momentum is somewhat noisy,
  as this represents a difficult limit for the \textsc{RNS} code
  \citep{Stergioulas95}; we could have used the slow-rotation
  approximation to compute the data in this limit, but we have preferred
  to perform the analysis on a single dataset.}. Each sequence terminates
with the ``maximum'' (critical) mass that is supported via (uniform)
rotation $M_{\rm max} := M_{\mathrm{crit}}(j=j_{\mathrm{Kep}})$, where
$j_{\rm Kep}$ is the maximum angular momentum that can be attained
normalised to the maximum mass, \ie $j_{\rm Kep} := J_{\rm
  Kep}/M^2_{\mathrm{Kep}}$. Note that the maximum mass in this case can
range from values as small as $M_{\rm max} \simeq 2.2 \,\mathrm{M}_{\bigodot}$ for $J
\simeq 4\,\mathrm{M}^2_{\bigodot}$, up to $M_{\rm max} \simeq 3.1 \,\mathrm{M}_{\bigodot}$ for $J
\simeq 5\,\mathrm{M}^2_{\bigodot}$.

The same data, however, can be expressed in terms of dimensionless
quantities and, more specifically, in terms of the critical mass
normalised to the maximum value of the corresponding nonrotating
configuration, \ie $\bar{M}:= M_{\rm crit}/M_{_{\rm TOV}}$ and of the
dimensionless angular momentum $j$ when the latter is normalized to the
maximum value allowed for that EOS, $j_{\rm Kep}$. Such a data is
collected in the right panel of Fig. \ref{fig:critical} and shows that
the variance in this case is extremely small. Indeed, it is possible to
express such a behaviour with a simple polynomial fitting function of the
type
\begin{equation} 
\label{eq:Mcrit}
\frac{M_{\mathrm{crit}}}{M_{_{\rm TOV}}}
=1 + a_{2}\left(\frac{j}{j_{\mathrm{Kep}}}\right)^{2} +
a_{4}\left(\frac{j}{j_{\mathrm{Kep}}}\right)^{4} \,,
\end{equation} 
where the coefficients are found to be $a_{2}=1.316\times 10^{-1}$ and
$a_{4}=7.111\times 10^{-2}$, with a reduced chi squared $\chi^{2}_{\rm
  red} = 3.586\times 10^{-5}$ and where, of course, $M_{\mathrm{crit}} =
M_{_{\rm TOV}}$ for $j=0$.

An immediate consequence of Eq. \eqref{eq:Mcrit} is also a very important
result. Irrespective of the EOS, in fact, the maximum mass that can be
supported through uniform rotation is simply obtained after setting
$j=j_{\mathrm{Kep}}$ and is therefore given by 
\begin{align} 
\label{eq:Mcrit_2}
M_{\rm max} := M_{\mathrm{crit}}(j=j_{\mathrm{Kep}}) = &
\left(1 + a_{2} + a_{4}\right) {M_{_{\rm TOV}}} \nonumber \\
\simeq & \left(1.203 \pm 0.022\right) {M_{_{\rm TOV}}} \,,
\end{align} 
where we have taken as error not the statistical one (which would be far
smaller) but the largest one shown in the comparison between the fit and
the data.

In summary, although different EOSs give substantially different maximum
masses and are able to reach substantially different angular momenta, the
maximum mass that can be supported at the mass-shedding limit is
essentially universal and is about 20\% larger than the corresponding
amount in the absence of rotation. Also shown in the bottom part of the
panel is the relative error between the normalised critical mass
$\bar{M}$ and its estimate coming from the fit \eqref{eq:Mcrit}, $M_{\rm
  fit}$. Note that the error for the largest angular momentum is below
$2\%$ for all the EOSs considered and that it is below $1\%$ for most
rotation rates.

Similar conclusions on the maximum mass were reached also by
\citet{Lasota1996}, although with a larger variance, probably due to the
use of EOSs that did not satisfy the constraint of $M_{_{\rm TOV}} >
2\,\mathrm{M}_{\bigodot}$ and different normalizations. This confirms the idea that
the uncertainties in correlations of this type could be further reduced
if future observations will increase the value of the maximum mass
$M_{_{\rm TOV}}$ \citep{Lattimer2015}. More importantly, expression
\eqref{eq:Mcrit} highlights that a universal relation is present not only
for the maximum critical mass, but is is valid also for any value of the
dimensionless angular momentum, with a variance that is actually smaller
for slowly rotating models. This is a result that was not discussed by
\citet{Lasota1996}.

Despite the important implications that expression \eqref{eq:Mcrit} has,
its use in the form above is very limited as it requires the implicit
knowledge of $M_{_{\rm crit}}$, as well as of $j$ and $j_{_{\rm Kep}}$. A
way round this limitation is possible if one makes the reasonable
assumption that for a given angular momentum, the mass and radius of the
mass-shedding model is proportional to the mass and radius of the
maximum-mass nonrotating model, \ie
\begin{align}
\label{eq:a1}
M_{\rm Kep} & \simeq b_1 M_{_{\rm TOV}}\,,&
R_{\rm Kep} & \simeq b_2 R_{_{\rm TOV}}\,,&
\end{align}
and that the Keplerian frequency follows the classical scaling in mass
and radius
\begin{align}
\label{eq:a3}
\Omega_{\rm Kep} & 
\simeq \sqrt{M_{\rm Kep}/R^{3}_{\rm Kep}}
\simeq b_3 \sqrt{M_{_{\rm TOV}}/R^{3}_{_{\rm TOV}}}\,.&
\end{align}
As a result, the angular momentum at the mass-shedding limit can be
expressed also as a function of the maximum-mass nonrotating
configuration
\begin{align}
\label{eq:a4}
J_{\mathrm{Kep}} &= I_{\rm Kep}\, \Omega_{\rm Kep} \simeq 
M_{\rm Kep}\, R^2_{\rm Kep}\, \Omega_{\rm Kep}
\simeq b_4 \sqrt{M^{3}_{_{\rm TOV}}\, R_{_{\rm TOV}}}\,, &
\end{align}
and thus 
\begin{align}
\label{eq:a5}
j_{\rm Kep} & = J_{\rm Kep}/M_{\rm Kep}^{2}\simeq
b_4 \sqrt{R_{_{\rm TOV}}/M_{_{\rm TOV}}} = 
b_4 \sqrt{1/\mathscr{C}_{_{\rm TOV}}}\,.& 
\end{align}
The coefficients $b_1, b_2$, and $b_3$ can be computed via a fit to the
data, yielding: $b_{1}=1.204\pm 0.002$, $b_{2}=1.310\pm 0.014$, and
$b_{3}=(6.303\pm 0.022)\times 10^{-1}$, while the coefficient $b_4$ can
be derived as $b_4 = b_1\,b^2_2\,b_3$. Alternatively, and more
accurately, $b_4$ can be computed via a direct fit of expression
\eqref{eq:a4}, which then yields $b_4=(5.543\pm 0.068)\times 10^{-1}$.

We can now use expression \eqref{eq:a5} in the dimensionless fit
\eqref{eq:Mcrit} to obtain
\begin{align}
\label{eq:Mcrit_3}
M_{\mathrm{crit}}= \left(1 + 
c_{2}\, \mathscr{C}_{_{\rm TOV}}\,j^{2} + 
c_{4}\, \mathscr{C}_{_{\rm TOV}}^{2}\,j^{4}\right)
{M_{_{\rm TOV}}}\,,
\end{align}
where the coefficients have numerical values $c_{2}=4.283\times 10^{-1}$
and $c_{4}=7.533\times 10^{-1}$. Expression has the desired features,
since the critical mass is now function only of the stellar angular
momentum $J$ and of the properties (mass and radius) of the maximum-mass
nonorotating configuration, $M_{_{\rm TOV}}$ and $R_{_{\rm TOV}}$. Note
that expression \eqref{eq:Mcrit_3} is not in an explicit form, since the
right-hand side still contains information on the critical mass via $j =
J/M^2_{\rm crit}$. However, a simple root-finding algorithm can be used
to obtain $M_{\rm crit}$ once $J$, $M_{_{\rm TOV}}$, and $R_{_{\rm TOV}}$
are specified.

\begin{figure*}
\begin{center}
\includegraphics[width=0.49\textwidth]{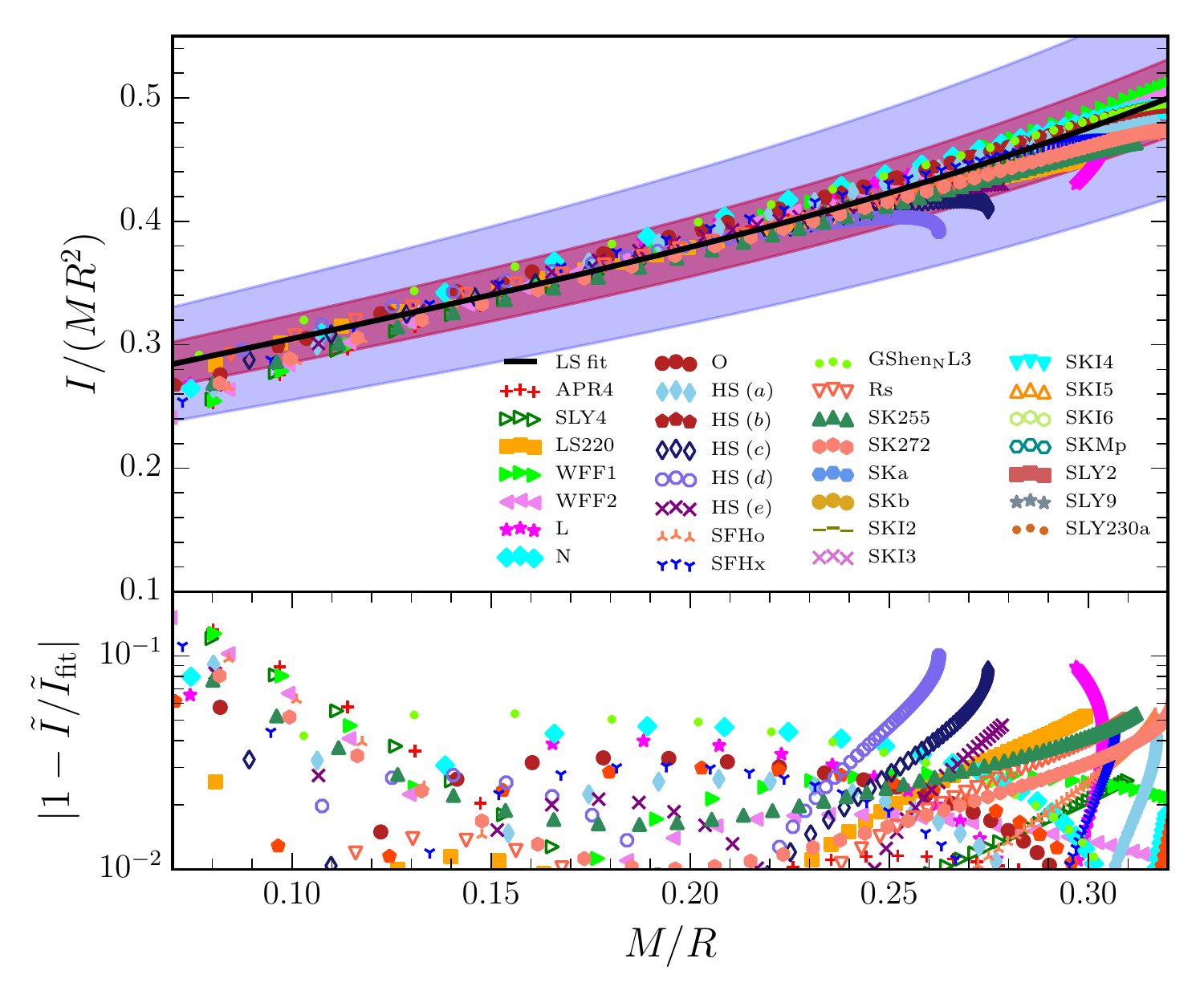}
\includegraphics[width=0.49\textwidth]{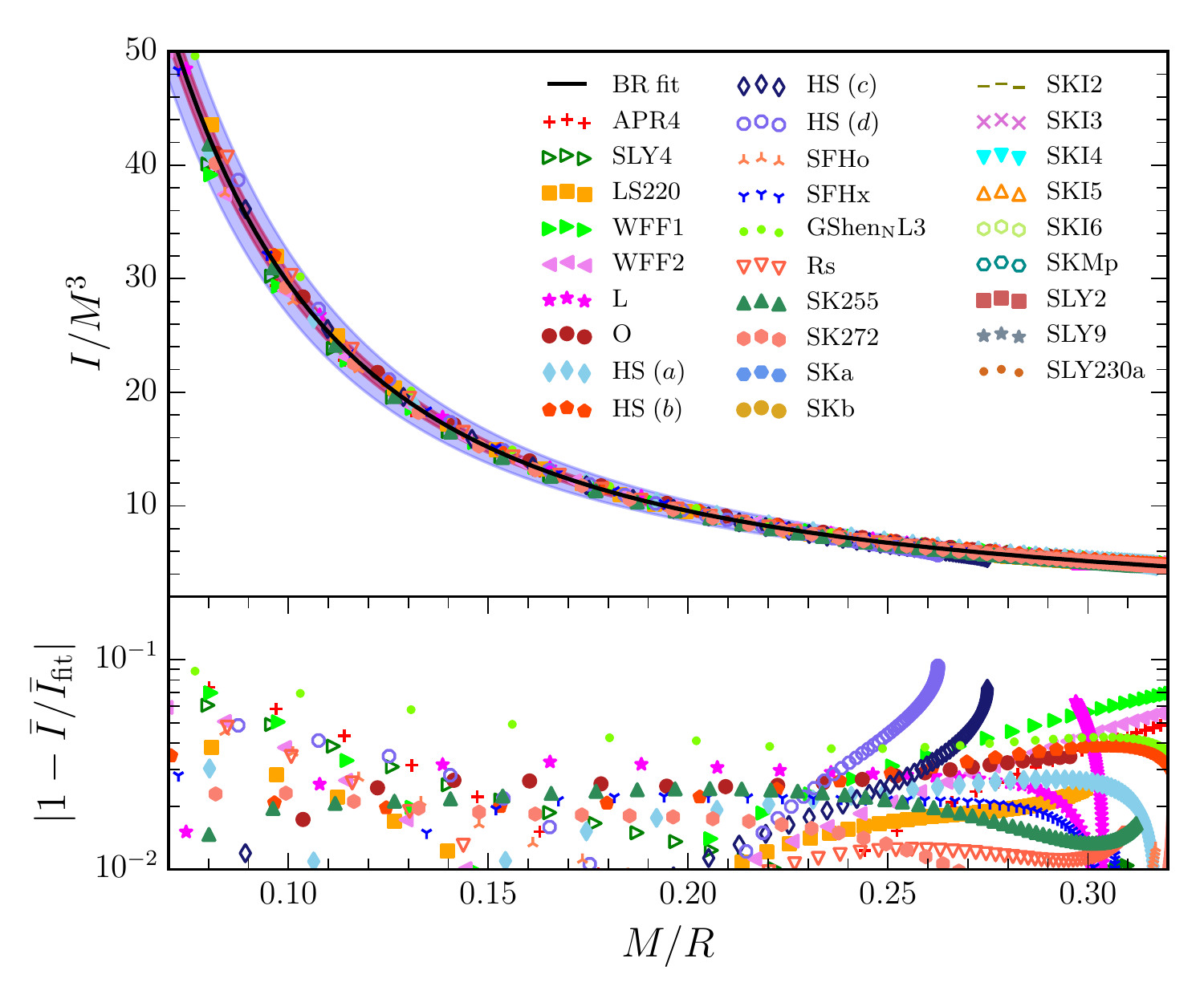}
\end{center}
\caption{\textit{Left panel:} Normalized moment of inertia $I/(MR^{2})$
  shown as a function of the stellar compactness $\mathscr{C}$, and for
  different EOSs, each indicated with a different symbol. The solid black
  line reports the fitting function (\ref{eq:LS_fit}) suggested by
  \citet{Lattimer2005b}. \textit{Right panel:} Normalized moment of
  inertia $I/M^{3}$ shown as a function of stellar compactness
  $\mathscr{C}$; also in this case, the solid black line reports the
  fitting function Eq. (\ref{eq:BR_fit}). In both panels the red- and
  blue-shaded areas refer respectively to the values of $\langle
  L_{\infty} \rangle$ and $L_{\infty}$, where $\langle L_{\infty}
  \rangle$ is the average over all EOSs of the largest residual $|1 -
  \bar{{I}} / \bar{{I}}_{\mathrm{fit}}|$, while $L_{\infty}$ is the
  largest residual across all EOSs (\cf Table \ref{table:norms}). Note
  that only $10\%$ of the points used in the analysis is shown in the
  plots.}
\label{fig:LS}
\end{figure*}

\section{$I$--$\mathscr{C}$ universal relations}
\label{sec:results}

In what follows we discuss universal relations between the moment of
inertia ${I}$ and the compactness $\mathscr{C}$. We will first consider
the case of slowly rotating stars (Section \ref{sec:IC_sr}) and then that
of rapidly rotating ones (Section \ref{sec:IC_rr}).

\subsection{$I$--$\mathscr{C}$ relations for slowly rotating stars}
\label{sec:IC_sr}

Starting with \citet{Ravenhall94}, a number of authors have pointed out
that the dimensionless moment of inertia $\tilde{I} := I/(MR^{2})$ can be
expressed in terms of the stellar compactness $\mathscr{C}$ via simple
low-order polynomial expressions that have a weak dependence on the
EOS. Expressions of this type have been proposed by \citet{Lattimer01},
as well as by \citet{Bejger02}, with comparable level of precision. A
higher-order polynomial fit was proposed subsequently by
\citet{Lattimer2005b} (LS), who expressed it as
\begin{equation}
\tilde{I} := 
\frac{I}{MR^{2}}=
\tilde{a}_{0}+\tilde{a}_{1}\,\mathscr{C}+
\tilde{a}_{4}\,\mathscr{C}^{4}\,,
\label{eq:LS_fit}
\end{equation}
and discussed how such a relation could be used to constrain the stellar
radius once $I$ and $M$ are measured, \eg in a pulsar within a binary
system. 

The left panel of Fig. \ref{fig:LS} reports with a scatter plot the
values of $\tilde{I}$ relative to slowly rotating stars [\ie stars
  treated in the slow-rotation approximation \eqref{eq:slowrot}] as
computed for a number of different EOSs and in a large range of
compactnesses, \ie with $\mathscr{C} \in [0.07, 0.32]$; this is also the
range in compactness over which the fits are performed. Different symbols
and colours refer to different EOSs as reported in the legend. Also
indicated with a black solid line is the LS fit expressed by
\eqref{eq:LS_fit}, while the shaded band band reports the error in
the fitting coefficients.

The fitting coefficients found originally by \citet{Lattimer2005b} are:
$\tilde{a}_{0}=0.237 \pm 0.008$, $\tilde{a}_{1}=0.674$ and
$\tilde{a}_{4}=4.48$, while those produced by our analysis are slightly
different and given by: $\tilde{a}_{0}=0.244$, $\tilde{a}_{1}=0.638$ and
$\tilde{a}_{4}=3.202$. The main differences in the estimates (especially
in the quartic term) are due to the different set of EOSs used and, more
importantly, to the fact that the estimates by \citet{Lattimer2005b} were
based in part on EOSs which are now excluded by observations of a 2 $\mathrm{M}_{bigodot}$ neutron star.
 
\citet{Lattimer01} and \citet{Lattimer2005b} found that fitting functions
of the type \eqref{eq:LS_fit} were valid for a wide range of EOS except
the ones which exhibited extreme softening. This is measured by the
relative error $|1-\tilde{I}/\tilde{I}_{\mathrm{fit}}|$, which is
reported in the lower part of the left panel of Fig. \ref{fig:LS}, and
which shows the errors are generally below $10\%$. At the same time, the
large scatter of the data suggests that a tighter fit could be obtained
if a different normalization is found for the moment of inertia. Hence,
inspired by the normalization's proposed by \citet{Lau2010} and
\citet{Yagi2013b}, we have considered to fit the moment of inertia
through a functional expansion of the type
\begin{equation}
\bar{I} := \frac{I}{M^{3}}=
    {\bar{a}_{1}}\,{\mathscr{C}^{-1}}+{\bar{a}_{2}}\,{\mathscr{C}^{-2}}+
    {\bar{a}_{3}}\,{\mathscr{C}^{-3}} +
    {\bar{a}_{4}}\,{\mathscr{C}^{-4}}\,.
\label{eq:BR_fit}
\end{equation} 
There are two main motivations behind this choice. The first one is that
it is clear that at lowest order $I/M^3 \sim 1/\mathscr{C}^2$ and hence
an expansion in terms of inverse powers of the compactness is rather
natural. The second one is the realisation that universal relations exist
between $I/M^3$ and $\lambda$ \citep{Yagi2013a} and between $\lambda$ and
$\mathscr{C}$ \citep{Maselli2013}, so that a universal relation should
also exist also between $I/M^3$ and $\mathscr{C}$, although not yet
discussed in the literature.

\begin{table}
\begin{center}
\begin{tabular}{l|cccc}
\hline\hline
           & $\langle L_1 \rangle$ & $\langle L_2 \rangle$ & $\langle L_\infty \rangle$ & $L_\infty$ \\
\hline
$I/(MR^2)$ & $0.01711$             & $0.00080$             & $0.06294$                & $0.16268$ \\
$I/M^3$    & $0.01068$             & $0.00043$             & $0.03209$                & $0.09420$ \\
\hline\hline
\end{tabular}
\caption{Summary of various averaged norms of the the residuals of the
  two fits, \ie the averages over all EOSs of all the residuals $|1 -
  \bar{{I}} / \bar{{I}}_{\mathrm{fit}}|$ when normalizing the moment of
  inertia as $I/(MR^2)$ or $I/M^3$, respectively. The last column reports
  instead the largest infinity norm, \ie the largest residual across all
  EOSs. The values of the last two columns are reported as shaded areas
  (red and blue, respectively) in Fig. \ref{fig:LS}.}
  \label{table:norms}
\end{center}
\end{table}

The right panel of Fig. \ref{fig:LS} shows the same data as in the left
panel but with the new normalisation $I/M^3$ and the new fitting function
\eqref{eq:BR_fit} with a black solid line, while the shaded grey band
band reports the error in the fitting coefficients. The behaviour of the
moment of inertia is now on average more accurately captured, with a
behaviour which is essentially universal and relative errors which are
comparable but also smaller than those of the fit \eqref{eq:LS_fit}\footnote{We
  note that since $I/(MR^2)$ and $I/M^3$ have intrinsically different
  magnitudes, the use of a $\chi^2$ would not be sensible for a
  comparison of the goodness of the fits.}. More precisely, the $\langle
L_1 \rangle$ norm of the residuals of the two fits, \ie the average over
all EOSs of all the residuals $|1 - \bar{{I}} /
\bar{{I}}_{\mathrm{fit}}|$) is given by $\langle L_1 \rangle \simeq
1.7\%$ for the fit of $I/(MR^2)$ and by $\langle L_1 \rangle \simeq 1.1\%
$ for the fit of $I/M^3$. Similar are also the values of the $\langle
L_\infty \rangle$ norms (\ie the largest relative deviation between the
data and the fits), which are $6\%$ and $3\%$, respectively. A summary of
the values of the various norms is shown in Table \ref{table:norms},
where the values of the last two columns are reported as shaded areas
(red and blue, respectively) in Fig. \ref{fig:LS}. Not surprisingly, the
largest errors are seen for very stiff (small $\mathscr{C}$) or very soft
(large $\mathscr{C}$) EOSs. The numerical values of the fitting
coefficients in Eq. \eqref{eq:BR_fit} are summarised in Table
\ref{table:fit_coeffs}, together with the corresponding reduced chi
squared.

\begin{table*}
\begin{tabular}{r|ccccc}
\hline\hline
                      & $\bar{a}_1$                     & $\bar{a}_2$                      & $\bar{a}_3$ & $\bar{a}_4$                       & $\chi^2_{\rm red}$\\
\hline
$\mathrm{slow\ rot.}$ & $8.134\times 10^{-1}$ & $2.101\times 10^{-1}$ & $3.175\times 10^{-3}$ & $-2.717\times 10^{-4}$ &$1.184\times 10^{-1}$\\\
$j=0.2$               & $9.426\times 10^{-1}$ & $1.541\times 10^{-1}$ & $1.117\times 10^{-2}$ & $-7.004\times 10^{-4}$ & $8.579\times 10^{-2}$\\
$j=0.4$               & $9.499\times 10^{-1}$ & $1.436\times 10^{-1}$ & $1.220\times 10^{-2}$ & $-7.613\times 10^{-4}$ & $1.047\times 10^{-1}$ \\
$j=0.6$               & $7.408\times 10^{-1}$ & $2.288\times 10^{-2}$ & $-2.975\times 10^{-3}$& $-9.895\times 10^{-5}$ & $1.655\times 10^{-1}$\\
\hline\hline
\end{tabular}
\caption{Summary of the fitting coefficients for the universal
  $\bar{I}$--$\mathscr{C}$ relation \eqref{eq:BR_fit} as obtained in the
  slow-rotation approximation or along sequences of constant
  dimensionless spin parameter. Also shown in the last column is the
  corresponding reduced chi squared.}
\label{table:fit_coeffs} 
\end{table*}

\begin{figure*}
\begin{center}
\includegraphics[width=0.49\textwidth]{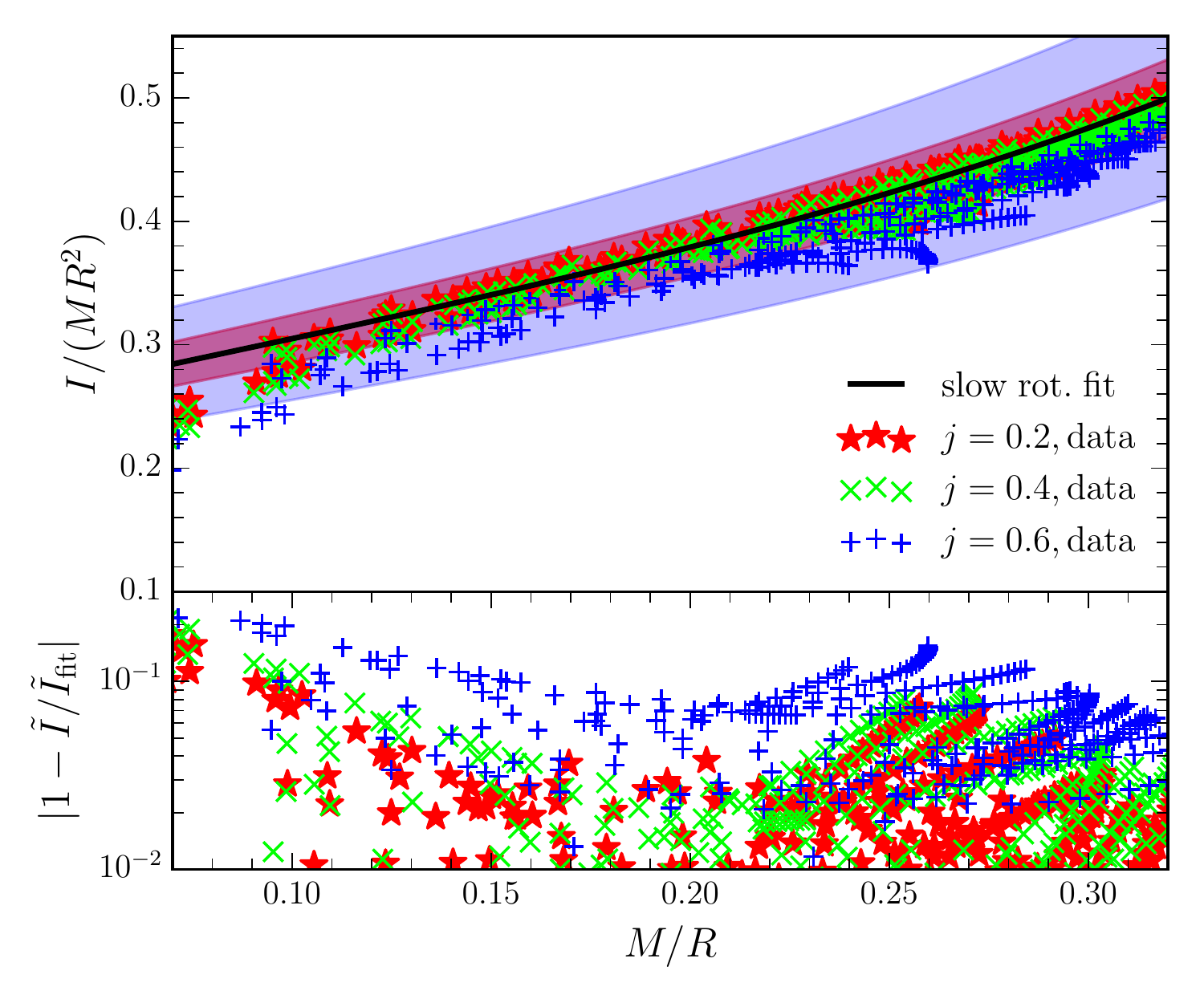}
\includegraphics[width=0.49\textwidth]{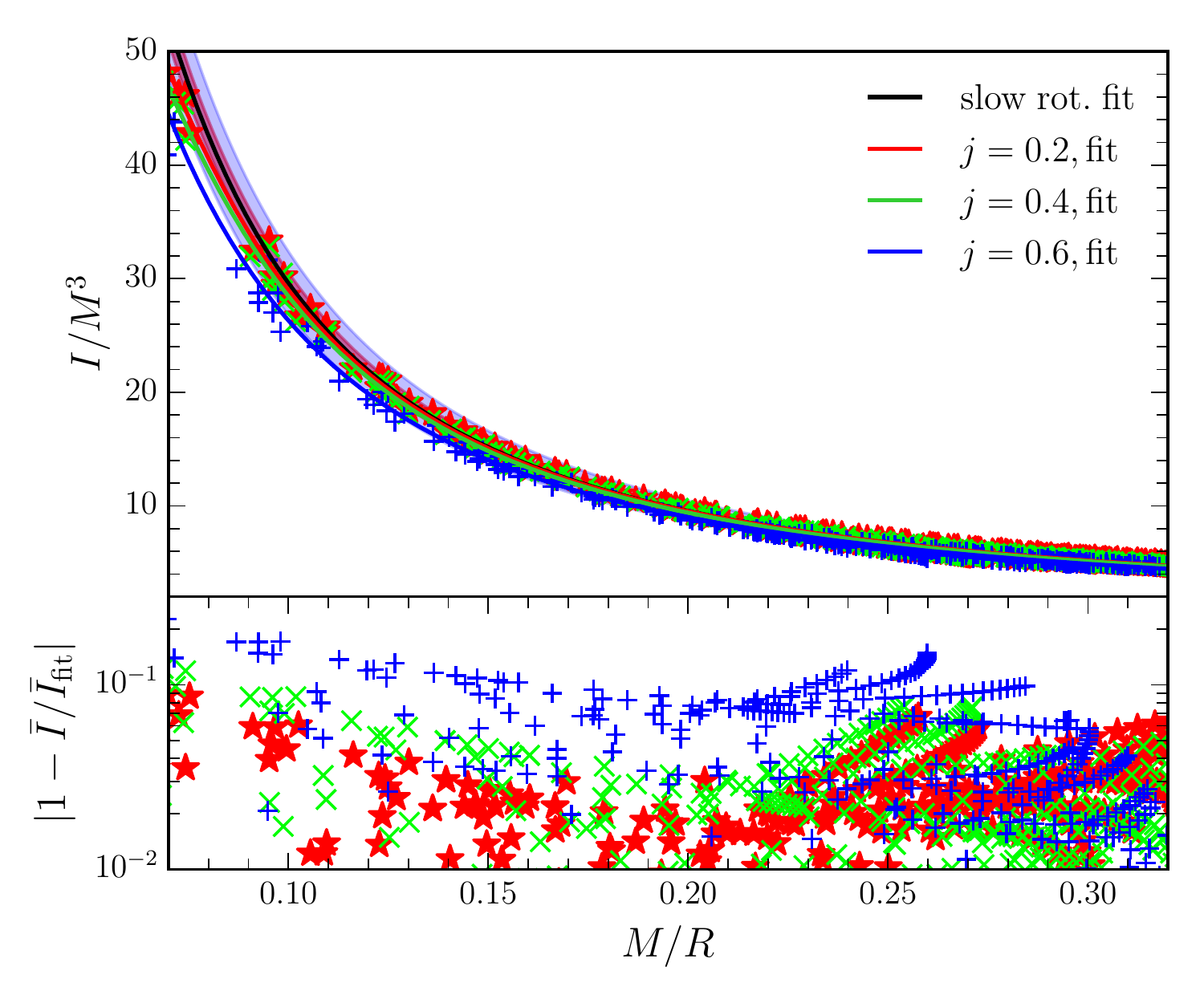}
\end{center}
\caption{\textit{Left panel:} Normalized moment of inertia $I/(MR^{2})$
  shown as a function of the stellar compactness $\mathscr{C}$. Different
  symbols refer to different EOSs, while symbols of a given colour refer
  to a certain value of the normalized angular momentum: $j=0.2$ (red
  symbols), $j=0.4$ (green symbols), and $j=0.6$ (blue symbols). The
  shaded area marks the uncertainty band of the fitting function for $I$
  in the slow-rotation approximation, while the bottom panel shows the
  relative error with respect to the fit (also in
  this case only $10\%$ of the points used in the analysis is shown in
  the plots).  \textit{Right panel:} The same as in the left panel but
  for the normalized moment of inertia $I/M^3$. In both panels, the light shaded
  areas refer to the values of the $L_{\infty}$ norms of the residuals for the
  slowly rotating stellar models, while the dark shaded areas mark the average of the $L_{\infty}$ norms over all EOSs. Overall, the data show that the
  universal behaviour found in the slow-rotation approximation is valid,
  within the errors, also for rapidly rotating stars.}
\label{fig:LSrot}
\end{figure*}

Three remarks are worth making. First, the fitting \eqref{eq:BR_fit}
could be further improved if the normalization of the moment of inertia
is modified in terms of dimensionless quantities, \eg a given power of
the compactness as suggested by \citet{Majumder2015}. While useful to
tighten the fit, we are not certain that this approach is effective in
general or provides additional insight; hence, we will not adopt it
here. Second, the quantity $\bar{I}$ is sometimes associated to what is
called the ``effective compactness'' $\eta := \sqrt{M^3/I} \sim M/R$
\citep{Lau2010}, where one is considering that $I \sim M R^2$. Clearly,
as the fit in expression \eqref{eq:BR_fit} reveals, this association is
reasonable only at the lowest order. Finally, while the relation
$\bar{I}$--$\mathscr{C}$ provides a description of the universal
behaviour of the moment of inertia that is slightly more accurate than
the one captured by the $\tilde{I}$--$\mathscr{C}$ relation, the visual
impression provided by Fig. \ref{fig:LS} enhances the impression of a
better fit for the $\bar{I}$--$\mathscr{C}$ relation. In practice, the
much larger range span by $\bar{I}$ across the relevant compactnesses
induces one to believe that the universality is far superior in this
latter case. As testified by the data in the lower plots, the error is,
on average, only slightly smaller.

\subsection{$I$--$\mathscr{C}$ relations for rapidly rotating stars}
\label{sec:IC_rr}

As mentioned in Section \ref{sec:setup}, within the slow-rotation
approximation the properties of the compact stars do not depend on the
magnitude of rotation and the surface is still given by a 2-sphere of
constant radial coordinate. When the stars are rapidly rotating, on the
other hand, not only the effect of the frame dragging needs to be
included, but also the change in the stellar surface needs to be taken
into account. More specifically, as a rotating star approaches the
``mass-shedding'' (or Keplerian) limit, that is, the limit at which it is
spinning so fast so as to lose matter at the equator, its mass and
equatorial radius increase while the polar radius decreases, leading to
an overall oblate shape. Since the moment of inertia is influenced mostly
by the mass in the outer regions of the star, the increases in mass and
radius lead quite generically to larger moments of inertia.

Since the $I$--$\lambda$--$Q$ universal relations were first discussed in
the limit of slow rotation and small tidal deformations,
\citet{Doneva2014a} have investigated the impact of rapid uniform
rotation on these relations and, in particular, on the between the moment
of inertia for sequences with constant angular velocity $\Omega$. What
was found in this case is that the $\bar{I}-Q$ universality was lost,
with deviations from the slow-rotation limit of up to 40\% with
increasing rotation rate. Interestingly, however, \citet{Chakrabarti2014}
also found that the universality is restored (within limits) if the
dimensionless moment of inertia is ordered not along sequences of
constant $\Omega$, but along sequences of constant dimensionless spin
parameter $j:=J/M^2$. In particular, along each sequence the $\bar{I}-Q$
relation is independent of the EOS up to about 1\%, but it is of course a
function of the spin parameter $j$.

Hence, it is natural to investigate whether the $\bar{I}-\mathscr{C}$
relation we have discussed in the previous Section in the case of slowly
rotating stars, continues to be valid also in the case of rapid
rotation. This is summarised in Fig. \ref{fig:LSrot}, whose two panels
show respectively $\tilde{I}$ and $\bar{I}$ as a function of
$\mathscr{C}$, with the shaded band marking the uncertainty band
coming from the fit within the slow-rotation approximation\footnote{To
  avoid overloading the figure, which has now four times the amount of
  data shown in Fig. \ref{fig:LS}, the data reported in
  Fig. \ref{fig:LSrot} refers only to the first EOSs in
  Fig. \ref{fig:LS}, \ie EOSs: APR4, Sly4, LS220, WFF1, WFF2, L, N, O, HS
  $(a)$, HS $(b)$, HS $(c)$, HS $(d)$, SFHo, SFHx.}. Following
\citet{Chakrabarti2014}, and expecting it will yield a tighter
correlation, we order the data along sequences of constant $j$. Note that
while the data for $j=0.2$ (red symbols) lies almost on top of the
results for the slow-rotation limit, the sequences for $j=0.4$
(light-green) show larger deviations, and for $j=0.6$ (blue symbols)
the data lies completely outside of the error band for the slow-rotation
fit. Indeed, along the $j=0.6$ sequence, which is also near the
mass-shedding limit for most EOSs, the deviation in the fitted range can
be as large as $20\%$.

Despite the complete loss of universality at high rotation rates and
small compactnesses, and the fact that universality is present only along
specific directions (\ie $j=\mathrm{const.}$ sequences), the
$\tilde{I}$--$\mathscr{C}$ relation remains effective for a large portion
of the space of parameters. We recall, in fact, that the fastest known
rotating pulsar has a spin frequency of 716 Hz \citep[or a period of
  $\simeq 1.4\,\mathrm{ms}$,][]{Hessels2006}, which could correspond to $j
\lesssim 0.5$ depending on the pulsar's mass \citep{Stein2014}. For the
EOSs used here, however, $j \lesssim 0.3$ for this pulsar and would
therefore lie within the error band of the fit for the slow-rotation
limit for most of that range. For completeness, and following what was done
for the slow-rotation approximation, we report in Table
\ref{table:fit_coeffs} the numerical values of the fitting coefficients
in Eq. \eqref{eq:BR_fit} for the various $j=\mathrm{const.}$ sequences
considered.

\begin{figure*}
\begin{center}
\includegraphics[width=0.49\textwidth]{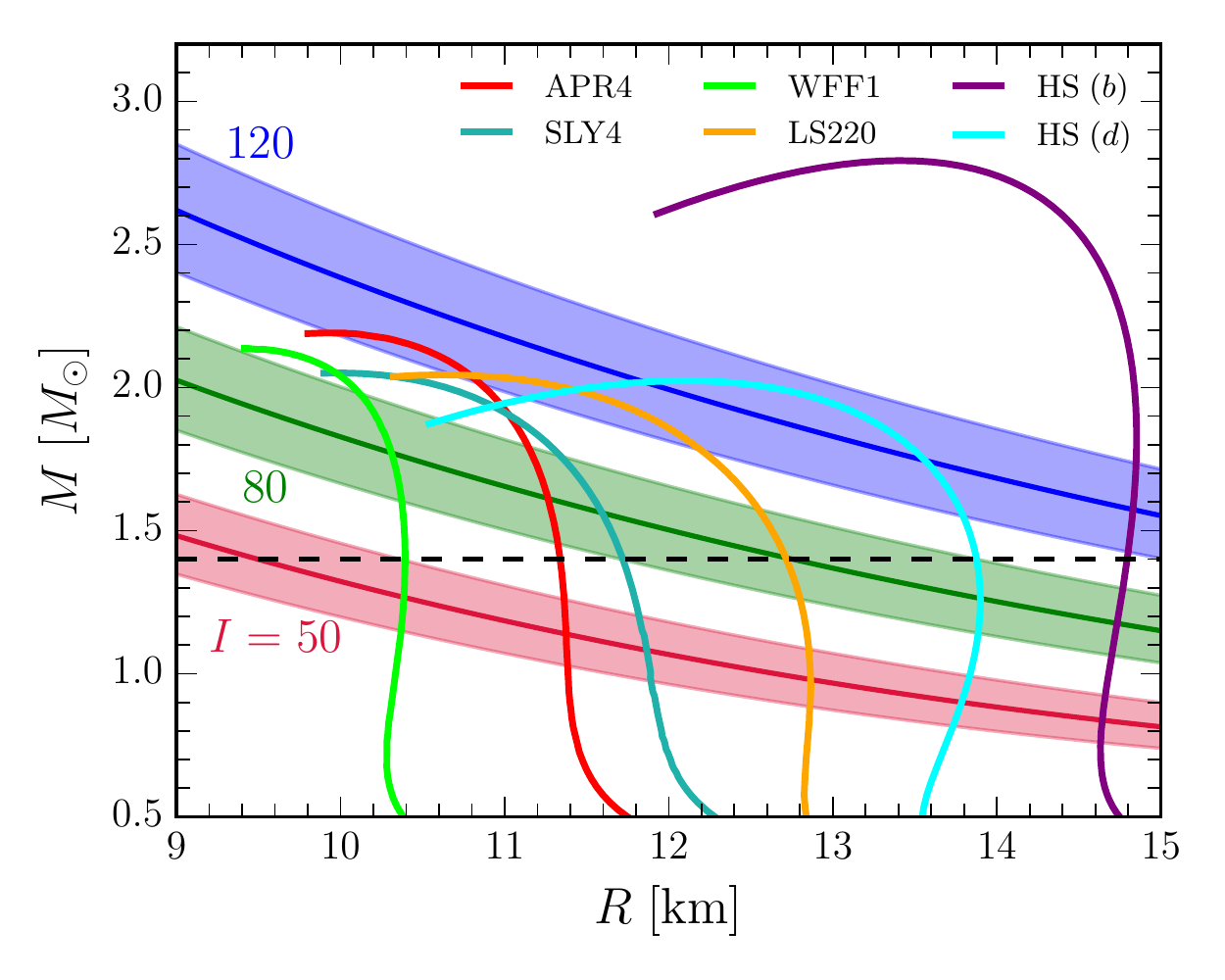}
\includegraphics[width=0.49\textwidth]{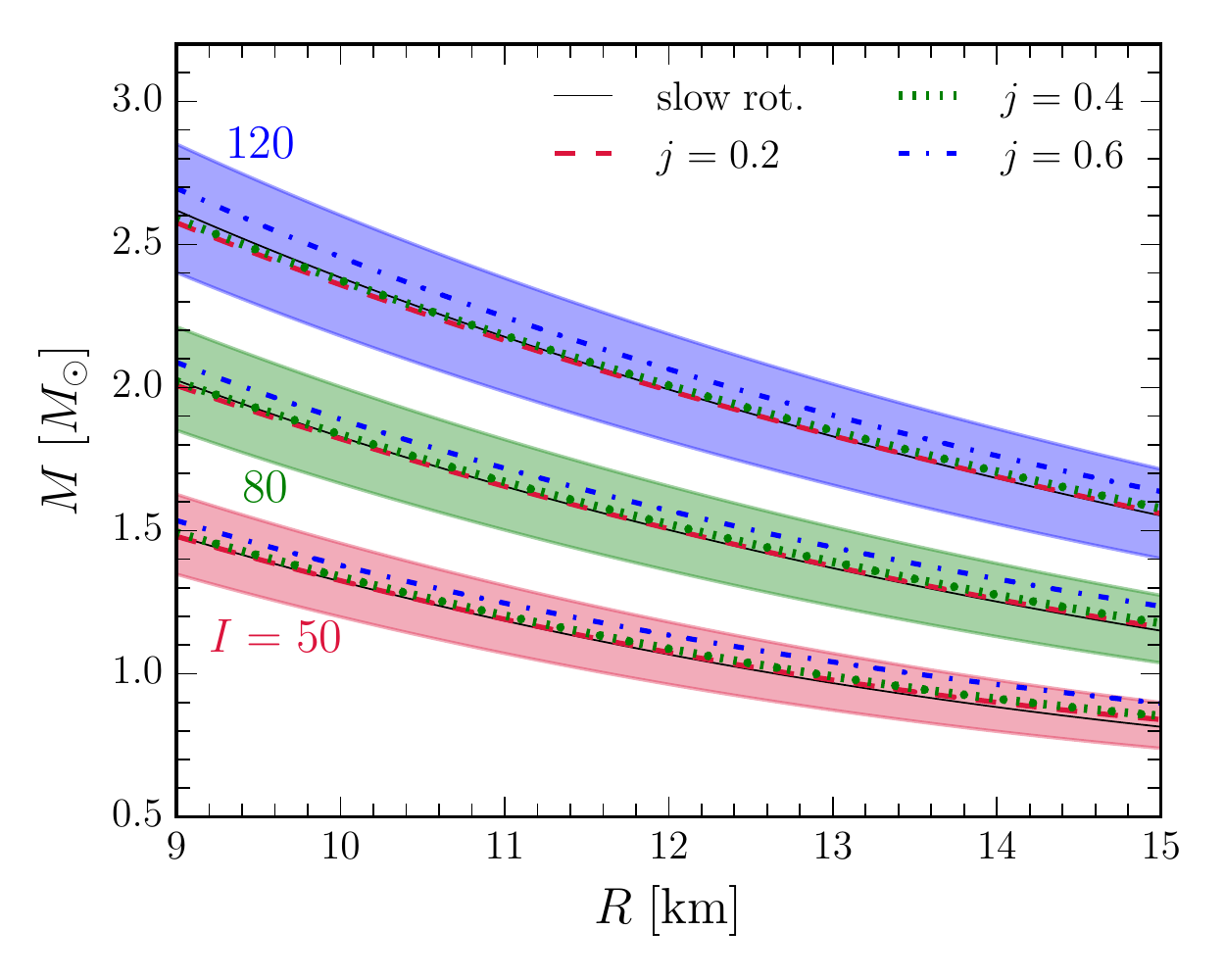}
\end{center}
\caption{\textit{Left panel:} Solid lines of different colours indicate
  the mass-radius relations for nonrotating stellar models for some
  representative EOSs. The red, green, and blue solid lines indicate
  instead the mass-radius relations as constrained by a measurement of
  the moment of inertia and through the inversion of
  Eq. (\ref{eq:BR_fit}). In particular, the red, green, and blue lines
  refer to moments of inertia of $I=50,\, 80$ and
  $120\,\mathrm{M}_{\bigodot}\,\mathrm{km}^{2}$, respectively. The corresponding
  shaded bands include an observational error on the moment of inertia of
  10\% and the uncertainty in the fitting coefficients. The black dashed
  line shows the specific case of a mass measurement of
  $M=1.4\,\mathrm{M}_{\bigodot}$. \textit{Right panel:} The same as in the left
  panel, but for rotating models organised in sequences of constant spin
  parameter $j$ (red, green, and blue dashed lines for $j=0.2,\, 0.4$ and
  $j=0.6$, respectively). Note that the curves for the rotating models
  lie within the error bands of the mass-radius relations computed for
  slowly rotating stars.}
\label{fig:radiusconstr}
\end{figure*}

\subsection{Universal relations with the binding energy}
\label{sec:BE}

Already \citet{Lattimer01} had pointed out that the binding energy of a
compact star, defined as the difference between the total rest mass
$M_{\rm b}$ and the gravitational mass of an equilibrium configuration,
${\rm BE} := M_{\rm b} - M$ could show a behaviour that is essentially
independent of the EOS. In particular, considering the dimensionless
binding energy ${\rm BE}/M$, \citet{Lattimer01} found a good fit to the
data with an expression of the type
\begin{equation}
\label{eq:BE_1}
\frac{{\rm BE}}{M} = \frac{c_1 \mathscr{C}}{1-c_2\mathscr{C}}\,,
\end{equation}
where $c_1=0.6$ and $c_2=0.5$. We have revisited the ansatz
\eqref{eq:BE_1} using more modern EOSs satisfying the two solar-mass
constraint and found that the fit to the data yields values for the
coefficients and in particular that $c_1=6.213\times 10^{-1} $ and
$c_2=1.941\times 10^{-1}$. The reduced chi squared for the two different
sets of coefficients is different, being $\chi^2_{\rm red} = 8.254\times
10^{-4}$ for the original fit of \citet{Lattimer01} and of $\chi^2_{\rm
  red} = 7.553\times 10^{-5}$ for our new fit.

Yet, because expression \eqref{eq:BE_3} is effectively a second-order
polynomial for small values of $\mathscr{C}$, we have also considered a
different functional form for the fitting ansatz and which, in the spirit
of expression \eqref{eq:BR_fit}, is a quadratic polynomial in the stellar
compactness, \ie
\begin{equation}
\label{eq:BE_2}
\frac{{\rm BE}}{M} = d_1 \mathscr{C} + d_2 \mathscr{C}^2\,,
\end{equation}
where $d_1= 6.19\times 10^{-1}$ and $d_2 = 1.359\times 10^{-1}$. The fit
in this case is marginally better than that obtained with expression
\eqref{eq:BE_2}, with a reduced chi square of $\chi^2_{\rm red} =
7.547\times 10^{-5}$. All things considered, and on the basis of the EOSs
used here, expression \eqref{eq:BE_2} is only a marginally better
description of the functional behaviour of the reduced binding energy and
should be considered effectively equivalent to expression
\eqref{eq:BE_1}.

Interestingly, expression \eqref{eq:BE_2} can also be easily extended to
encompass the case of rotating stars after replacing the coefficients
$d_1$ and $d_2$ with new expressions that contain rotation-induced
corrections in terms of the dimensionless angular momentum, \ie $d_1 \to
\tilde{d}_1$ and $d_2 \to \tilde{d}_2$ in \eqref{eq:BE_2}, where 
\begin{equation}
\label{eq:BE_3}
\tilde{d}_1 := d_1 (1+ \alpha j + \beta j^2)\,, \qquad
\tilde{d}_2 := d_2 (1+ \gamma j + \delta j^2)\,,
\end{equation}
where $\alpha = -1.966\times 10^{-2}$, $\beta = 4.272\times 10^{-1}$,
$\delta = -7.603$, and $\gamma = 4.46\times 10^{-1}$. Note that the
linear coefficients are very small, as are the errors in the estimates of
the quadratic coefficients.

The reduced chi square obtained when comparing the values of the reduced
binding energy ${\rm BE}/M$ obtained via \eqref{eq:BE_2}--\eqref{eq:BE_3}
with those obtained from the RNS is rather small and given by
$\chi^2_{\rm red} = 1.624\times 10^{-4}$. Finally, we note that
expressing the binding energy in terms of the radius rather than of the
mass changes the functional behaviour of the fitting function since ${\rm
  BE}/R \sim \mathscr{C}^2$, but not the overall quality of the fit.

\section{Applications of the universal relation}
\label{sec:applications}

While the origin of the universal relations is still unclear and the
subject of an intense research (and debate), we take here the more
pragmatic view in which universal relations are seen as an interesting
behaviour of compact stars in a special area of the space of solutions:
namely, those of slowly rotating, low-magnetisation stars. In this view,
universal relations can be used to constrain phenomenologically
quantities which are not directly accessible by observations or whose
behaviour is degenerate. For instance, the simultaneous measurement of
the mass and of the moment of inertia of a compact star, \eg of a pulsar
in a binary system of compact stars, does not necessarily provide
information on the radius. On the other hand, the same measurements,
together with the use of the universal relation (\ref{eq:BR_fit}) can set
constraints on the radius and hence on the EOS. To illustrate how this
can be done in practice let us consider the case in which the moment of
inertia is measured with an observational error of 10\%; this may well
be an optimistic expectation but is not totally unrealistic [a similar
  assumption was made also by \citet{Lattimer2005b}]. If the mass is
observed with a much larger precision [as it is natural to expect from
  pulsar measurements in binary neutron-star systems~\citep{Kramer2009}],
then it is possible to define regions in the $(M,\,R)$ plane that are
compatible with such observations. In turn, the comparison with the
expectations from different EOSs will set constraints on the radius of
the star.

In practice, given a measurements of $I$ and $M$, it is sufficient to
invert the fitting function (\ref{eq:BR_fit}) to obtain the range of
radii that is compatible with the measurements of $I$ and $M$. Examples
of these compatibility regions in the $(M,\,R)$ plane are represented by
the coloured shaded regions in the left panel of
Fig. \ref{fig:radiusconstr}, where we have considered respectively
$I=50\,\mathrm{M}_{\bigodot}\,\mathrm{km^2}$ (red-shaded region),
$I=80\,\mathrm{M}_{\bigodot}\,\mathrm{km^2}$ (green-shaded region), and
$I=120\,\mathrm{M}_{\bigodot}\,\mathrm{km^2}$ (blue-shaded region)\footnote{Note that
  the shaded bands include both an observational error on the moment of
  inertia of 10\% and the error on the fit coming from
  Eq. (\ref{eq:BR_fit}).}. Given then a measure of the mass (we have
considered a canonical mass of $M=1.4\,\mathrm{M}_{\bigodot}$ and indicated it with
an horizontal black dashed line), the intersection of the corresponding
constraint with a given shaded region sets a range for the possible value
of the radius. For instance, a measurement of
$I=80\,\mathrm{M}_{\bigodot}\,\mathrm{km^2}$ and $M=1.4\,\mathrm{M}_{\bigodot}$ would require
radii in excess of $\simeq 12\,\mathrm{km}$, thus excluding relatively
soft EOSs such as APR4.

Similar considerations can be made also for rapidly rotating stars. This
is shown as is shown in the right-hand panel of Fig. \ref{fig:radiusconstr},
where we consider again three different hypothetical measurements of the
moment of inertia, but also consider the extension of the universal
relation \eqref{eq:BR_fit} along sequences of constant dimensionless spin
parameter $j$. We indicate with a black solid line the fit relative to
the slowly rotating models and with red, green and blue dashed lines the
fits corresponding to $j=0.2,\,0.4$, and $j=0.6$, respectively. Note that
even the largest considered value for $j$, the sequences are within the
error bands imposed by the observational error on $I$ and by the
uncertainties on the fitting function.

Following the considerations above, we next compute the error made in
estimating the radius of the star for any EOS once a measurement is made
of $I$ and $M$. To this scope we need to account both for the error due
to the observational uncertainty of the moment of inertia and for the
error on the fitting function. Because the error on the fit
(\ref{eq:BR_fit}) is of the order of 5\% for a large part of the
considered range in the compactness, the error on the radius made when
inverting the fitting function (\ref{eq:BR_fit}) can be estimated via the
standard Gaussian error-propagation law. More specifically, we assume
that the total error in the radius estimate is given by the sum in
quadrature of two errors, \ie
\begin{equation}
\sigma_{_{\rm R}}=\sqrt{\sigma_{\rm fit}^{2} + 
\sigma_{I_{\rm obs}}^{2}}\,,
\end{equation}
where $\sigma_{\rm fit}$ is the error in the fit when computed as the
$L_{\infty}$-norm of the relative deviation of the fit from the data [\cf
  Eq. \eqref{eq:sigmafit}], while $\sigma_{I_{\rm obs}}$ is the error in
the observed moment of inertia. The error on the fit is approximately
given by
\begin{align}
\label{eq:sigmafit}
\sigma_{\rm fit}^{2}=\bar{I}^{2}L_{\infty}^{2}\,,
\end{align}
where $L_{\infty}$ is the maximum relative error $(1 -
\bar{I})/\bar{I}$. Of course, all what was discussed so far can be
applied equally to the new fitting relation \eqref{eq:BR_fit} as to the
original fit by \citet{Lattimer2005b} given by expression
\eqref{eq:LS_fit}.

\begin{figure}
\begin{center} 
\includegraphics[width=0.99\columnwidth]{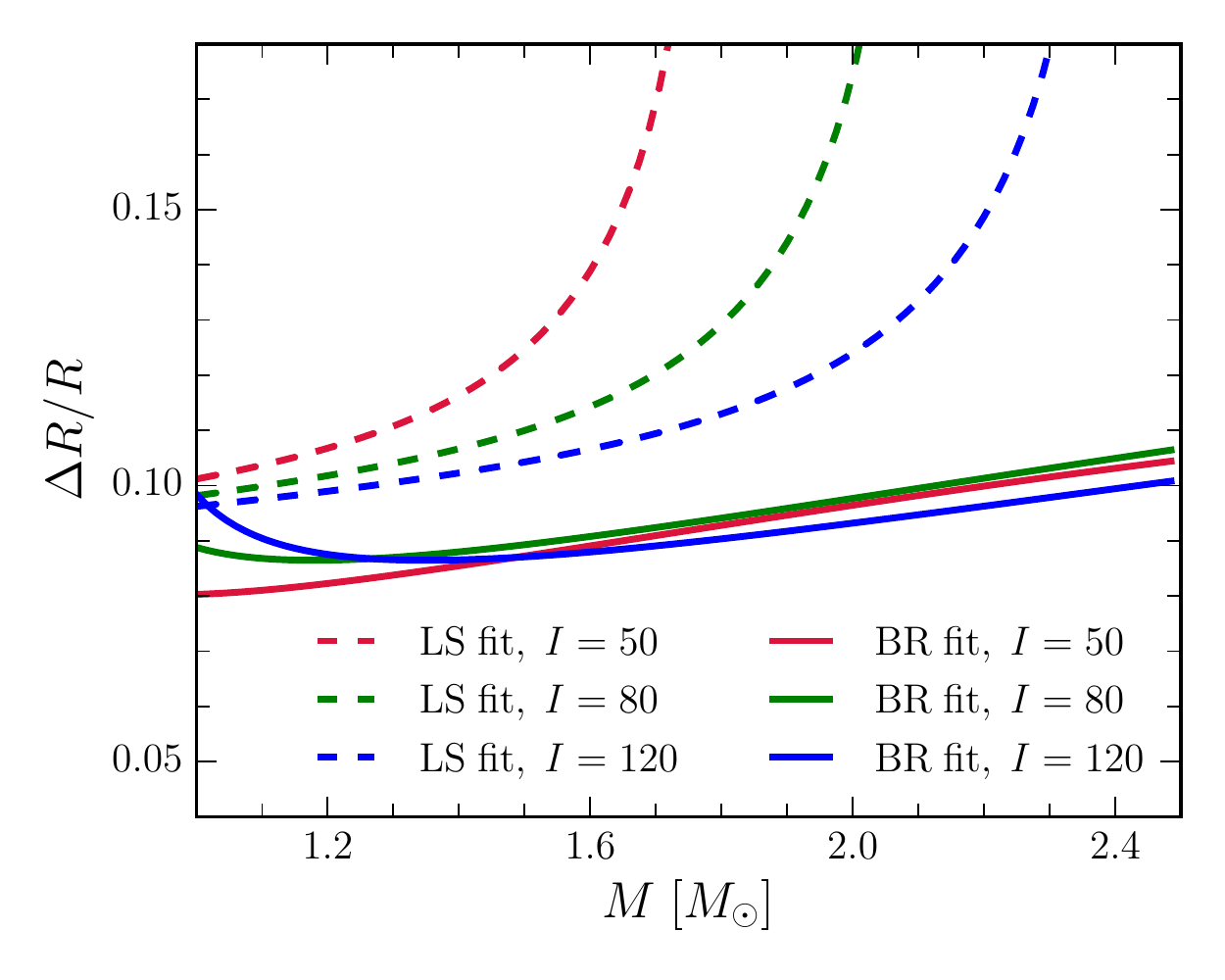}
\end{center} 
\caption{Relative error on the radius obtained either from the fit
  (\ref{eq:LS_fit}) (LS fit, dashed lines), or from the fit
  (\ref{eq:BR_fit}) (BR fit, solid lines). Different colours refer to
  different values of the moment of inertia (red, green and blue lines
  for $I=50,\,80$ and $120\,\mathrm{M}_{\bigodot}\, {\rm km}^2$, respectively). The
  error estimates includes an observational uncertainty of the moment of
  inertia of 10\% and the error on the fitting function.}
\label{fig:err}
\end{figure}

Figure \ref{fig:err} summarises the results of this error analysis by
reporting the relative error in the measurement of the stellar radius
$\Delta R/R$ for any EOS once the mass of the star is measured to high
precision and the moment of inertia is estimated with a relative error of
$10\%$. More specifically, Eqs. \eqref{eq:LS_fit} and \eqref{eq:BR_fit}
are inverted numerically to estimate $R$ from $M$ and $I$, while $\Delta
R$ is computed as the difference between the radius obtained from the
inverted fitting function and the inverted relation with the errors on
the moment of inertia and the fitting function taken into account.  

Figure \ref{fig:err} shows this relative error $\Delta R/R$ as a function
of the measured mass and also in this case we report with lines of
different colours the three different hypothetical measurements of the
moment of inertia. Note that we indicate with solid lines the errors
deduced using the new fit \eqref{eq:BR_fit} (BR fit) and with dashed
lines the corresponding fit coming from expression \eqref{eq:LS_fit} of
\citet{Lattimer2005b} (LS fit).

The two prescriptions provide rather different behaviours in the error
estimates as a function of mass. In particular, for small masses, \ie
$M\lesssim 1.4\,\mathrm{M}_{\bigodot}$, the modelling in terms of $\tilde{I}$ and
$\bar{I}$ yield comparable errors . However, for large masses, \ie
$M\gtrsim 1.4\,\mathrm{M}_{\bigodot}$, the modelling in terms of $\bar{I}$ yields
considerably smaller errors.

\section{Conclusions} 
\label{sec:conclusions}

We have shown that a universal relation is exhibited also by equilibrium
solutions that are not stable. In particular, we have considered
uniformly rotating configurations on the turning-point line, that is,
whose mass is an extremum along a sequence of constant angular
momentum. Such stellar configurations are unstable since they are found
at larger central rest-mass densities than those that on the
neutral-point line and therefore marginally stable
\citep{Takami:2011}. While hints of this relation were already discussed
in the literature \citep{Lasota1996}, we have here shown that it holds
not only for the maximum value of the angular momentum, but also for any
rotation rate. The importance of this universal relation is that it has
allowed us to compute the maximum mass sustainable through rapid uniform
rotation, finding that for any EOS it is about 20\% larger than the
maximum mass supported by the corresponding nonrotating configuration.

Finally, using universal relations for some of the properties of compact
stars, we have revisited the possibility of constraining the radius of a
compact star from the combined measurement of its mass and moment of
inertia, as is expected to be possible in a binary system containing a
pulsar. In particular, after considering both stellar models in the
slow-rotation approximation and in very rapid rotation, we have shown
that the dimensionless moment of inertia $\bar{I}=I/M^3$ for slowly
rotating compact stars correlates tightly and universally with the
stellar compactness $\mathscr{C}=M/R$ in a manner that does not depend on
the EOS. We have also derived an analytical expression for such a
correlation, which improves on a previous expression obtained with the
dimensionless moment of inertia $\tilde{I}=I/(MR^2)$
\citep{Lattimer2005b}.

Assuming that a measurement of the moment of inertia is made with a
realistic precision of 10\% and that a much more accurate measurement is
made of the mass, we have found that the new relation yields relative
errors in the estimate of the stellar radius that are $\lesssim 7\%$ for
a large range of masses and moment of inertia. Radius measurements of
this precision have the potential of setting serious constraints on the
EOS. Interestingly, the universal relation between $\bar{I}$ and
$\mathscr{C}$ is not restricted to slowly rotating models, but can be
found also in stellar models that are spinning near the mass-shedding
limit. In this case, the universal relation needs to be parameterized in
terms of the dimensionless angular momentum, but the functional behaviour
is very close to the nonrotating limit also at the most extreme rotation
rates.

\section{Acknowledgements} 
\label{sec:acknowledgements}

We thank V. Ferrari, B. Haskell, and J. Lattimer for useful discussions
and comments. We are also grateful to J. C. Miller for a careful read of
the paper, to S. Typel for essential help with the use of CompOSE, and to the
referee for useful comments and suggestions. LR thanks the Department of
Physics of the University of Oxford, where part of this work was carried
out. Partial support comes also from ``NewCompStar'', COST Action MP1304,
and by the LOEWE-Program in HIC for FAIR. All of the EOSs used here can
be found at the EOS repository ``CompStar Online Supernovae Equations of
State (CompOSE)'', at the URL \url{compose.obspm.fr}.

\section{Note added in Proof} 

Since the posting of this paper, a number of related works have been
published or posted. \citet{Bhattacharyya2016} have considered rapidly
rotating strange stars and the data in their tables suggest that a
universal maximum mass is present also for the quark-matter EOSs
considered (i.e., $M_{_{\rm max}} \sim 1.44\,M_{_{\rm TOV}}$), but
additional work is needed to confirm this behaviour. \citet{Staykov2016}
have instead considered compact stars in scalar-tensor theories and
${R}^2$ gravity, finding that relations similar to the ones reported here
are valid also in such theories.

\input{breu_rezzolla.bbl}

\appendix
\section{Analytical estimation of the error on the radius}

As discussed in Sec. \ref{sec:applications}, we reported in
Fig. \ref{fig:err} the relative error of the radius estimate, $\Delta
R/R$, as computed numerically for the LS fit (dashed lines) and for the
BR fit (continuous lines). As first pointed out by the referee, an
analytical expression for the LS fit can also be obtained under a number
of simplifying assumptions. To obtain such an expression, we first
express the moment of inertia as [\cf Eq. \eqref{eq:LS_fit}]
\begin{equation}
I = MR^2 \left( \tilde{a}_{0} + \tilde{a}_{1}\mathscr{C} + \tilde{a}_{4}
\mathscr{C}^4\right ) = MR^2 \gamma\,,
\end{equation}
so that $I=I(M, R, \tilde{a}_{0}, \tilde{a}_{1}, \tilde{a}_{4}) =
I(x^j)$, where we have indicated with $x^j$ the various degrees of
freedom. The error in the measure of $I$ can be expressed as
\begin{equation}
\Delta I = \left[ 
\sum_j
\left(\frac{\partial I}{\partial x^j}\right)^2
\left(\Delta x^j\right)^2
\right ]^{1/2}\,.
\end{equation}
Assuming $\Delta M = 0 = \Delta \tilde{a}_{1} = \Delta \tilde{a}_{4}$, as
in \citet{Lattimer2005b}, the error in the moment of inertia is
\begin{equation}
\label{eq:delta_I_0}
\Delta I = \left[ 
\left(\frac{\partial I}{\partial \tilde{a}_{0}}\right)^2
\left(\Delta \tilde{a}_{0}\right)^2
+
\left(\frac{\partial I}{\partial R}\right)^2
\left(\Delta R\right)^2
\right]^{1/2}\,.
\end{equation}

We can evaluate the two terms in \eqref{eq:delta_I_0} and obtain
\begin{equation}
\label{eq:delta_I_1}
\frac{\partial I}{\partial \tilde{a}_{0}} = MR^2 = \frac{I}{\gamma}\,,
\end{equation}
and 
\begin{equation}
\label{eq:delta_I_2}
\frac{\partial I}{\partial R} = 
M R\left(
2\gamma +R \frac{\partial \gamma}{\partial R} 
\right)
\,.
\end{equation}
A bit of algebra leads to 
\begin{equation}
\frac{\partial \gamma}{\partial R}  = 
-\frac{\mathscr{C}}{R}
\left(\tilde{a}_{1} + 4 \tilde{a}_{4} \mathscr{C}^3
\right)
\,,
\end{equation}
so that expression \eqref{eq:delta_I_2} can be written as 
\begin{equation}
\label{eq:delta_I_3}
\frac{\partial I}{\partial R} = 
MR\left(
2 \tilde{a}_{0} + \tilde{a}_{1} \mathscr{C} - 2 \tilde{a}_{4} \mathscr{C}^4
\right)
=
\tilde{a}_{0} M R \tilde{\chi}\,,
\end{equation}
where 
\begin{equation}
\tilde{\chi} := 2 + \frac{\tilde{a}_{1}}{\tilde{a}_{0}}\mathscr{C} - 
\frac{2 \tilde{a}_{4}}{\tilde{a}_{0}}\mathscr{C}^4\,.
\end{equation}

Using now \eqref{eq:delta_I_1} and \eqref{eq:delta_I_3}, we rewrite
\eqref{eq:delta_I_0} as 
\begin{equation}
\label{eq:delta_I_4}
\Delta I = 
\tilde{a}_{0} M R^2 
\left[ 
\left(\frac{\Delta \tilde{a}_{0}}{\tilde{a}_{0}}\right)^2
+
\tilde{\chi}^2
\left(\frac{\Delta R}{R}\right)^2
\right]^{1/2}\,.
\end{equation}
Recognising now that the first term on the right-hand side
of \eqref{eq:delta_I_4} is
\begin{equation}
\tilde{a}_{0} M R^2 = \frac{\tilde{a}_{0} I}{\gamma} = \frac{I}{\chi}\,,
\end{equation}
or, equivalently, that 
\begin{equation}
\chi = 1 + \frac{\tilde{a}_{1}}{\tilde{a}_{0}}\mathscr{C} +
\frac{\tilde{a}_{4}}{{\tilde a}_{0}}\mathscr{C}^4\,,
\end{equation}
can rewrite it as
\begin{equation}
\label{eq:delta_I_5}
\frac{\Delta I}{I} = 
\frac{1}{\chi}
\left[ 
\left(\frac{\Delta \tilde{a}_{0}}{\tilde{a}_{0}}\right)^2
+
\tilde{\chi}^2
\left(\frac{\Delta R}{R}\right)^2
\right]^{1/2}\,,
\end{equation}
so that
\begin{equation}
\label{eq:delta_I_6}
\frac{\Delta R}{R} = 
\frac{1}{\tilde{\chi}}
\left[ 
\chi^2\left(\frac{\Delta I}{I}\right)^2
-
\left(\frac{\Delta \tilde{a}_{0}}{\tilde{a}_{0}}\right)^2
\right]^{1/2}\,.
\end{equation}

We can next assume that the fitting error in the $\tilde{a}_{0}$ is also
very small, so that ${\Delta \tilde{a}_{0}}/{\tilde{a}_{0}} \sim 0$ and
thus Eq. \eqref{eq:delta_I_6} can be simply written as
\begin{equation}
\label{eq:delta_I_7}
\frac{\Delta R}{R} \sim 
\frac{\chi}{\tilde{\chi}}
\frac{\Delta I}{I}
= f(\mathscr{C})
\frac{\Delta I}{I}
\,,
\end{equation}
where the function $f(\mathscr{C})$ is defined as
\begin{equation}
\label{eq:delta_I_8}
f(\mathscr{C}) := \frac{\chi}{\tilde{\chi}} = \frac{1 +
  (\tilde{a}_{1}/\tilde{a}_{0})\mathscr{C} + (\tilde{a}_{4}/{\tilde
    a}_{0})\mathscr{C}^4} {2 + (\tilde{a}_{1}/\tilde{a}_{0})\mathscr{C} -
  2(\tilde{a}_{4}/\tilde{a}_{0})\mathscr{C}^4} \,,
\end{equation}
so that, once a value for the moment of inertia is fixed, expressions
\eqref{eq:LS_fit} and \eqref{eq:delta_I_8} can be written as a function
of the mass $M$ only. A similar procedure can be carried out for the
fitting in Eq. \eqref{eq:BR_fit}.

The behaviour of the denominator in the function $f(\mathscr{C})$ shows
that the error can diverge for a given value of the compactness. For
Eq. \eqref{eq:delta_I_8}, which is only a first approximation since it
neglects the error in the moment of inertia due to the spread in the
EOSs, this divergence occurs at $\mathscr{C} \sim 0.554$ when using the
coefficients in Eq. \eqref{eq:LS_fit}. However, in the fully numerical
analysis carried out in Fig. \ref{fig:err}, this divergence takes place
at smaller compactnesses and around $\mathscr{C} \sim 0.255$. We also
note that because $f(\mathscr{C})$ depends on the compactness, there will
be different relative errors $\Delta R/R$ for different choices of the
moment of inertia, as shown in Fig. \ref{fig:err}.

\label{lastpage}
\end{document}